\documentclass[aps,preprint]{revtex4}%
\usepackage{amsfonts}
\usepackage{amsmath}
\usepackage{amssymb}
\usepackage{graphicx}%
\begin{document}
\title[ ]{A\ modified N=2 extended supersymmetry}
\author{N. Djeghloul}
\affiliation{Laboratoire de Physique Th\'{e}orique d'Oran (LPTO), Universit\'{e} d'Oran, Alg\'{e}rie.}
\author{M. Tahiri}
\affiliation{Laboratoire de Physique Th\'{e}orique d'Oran (LPTO), Universit\'{e} d'Oran, Alg\'{e}rie.}


\begin{abstract}
A modification of the usual extended $N=2$ supersymmetry algebra implementing
the two dimensional permutation group is performed. It is shown that one can
found a multiplet that forms an off-shell realization of this alternative
extension of standard supersymmetry.


Shell document for REV\TeX{} 4.

\end{abstract}
\startpage{01}
\endpage{02}
\maketitle

\section{\bigskip Introduction}

\bigskip The present paper deals with the possibility of modifying the usual
extended $N=2$ super Poincar\'{e} algebra via a suitable implementation of the
symmetric group $S_{2}$. This construction has the interesting advantage that
one can find a multiplet that realizes the modified extended supersymmetry
algebra \textit{off-shell}. This multiplet involves the same fields as those
of the standard double tensor multiplet \cite{1} (see also \cite{2} for
explicit construction). It is worth mentioning that this construction is not a
standard extension of supersymmetry in the sense that it rely on a non-local
invariance represented by the symmetric group $S_{2}$. This is what the term
\textit{modified} underlies. The obtained transformations still transform
bosons in fermions and vice-versa.

We will first show that a suitable modification of the $N=2$ supersymmetry
algebra is possible in the context of nonlinear extension of standard Lie
algebras \cite{3}. In this context, we introduce the symmetric group $S_{2}$
within the standard extended $N=2$ super Poincar\'{e} algebra \cite{4}. This
is what is explicitly performed in section two.

In section three, we show that the multiplet containing two Weyl fermions, two
real scalar fields and two $2-$form gauge potentials is an off-shell multiplet
of the modified superalgebra so that no auxiliary fields are needed. Finally
we show that the construction of a nilpotent BRST operator can be considered.

\section{Modified N=2 Supersymmetry algebra}

\bigskip The possibility of modifying the extended $N=2$ super Poincar\'{e}
algebra is based on the observation that the free Lagrangian density and thus
the free action of two scalar fields $\varphi_{i}$ $(i=1,2)$ or two spinor
fields $\psi^{a}$ $(a=1,2)$ (hereafter a repeated index means a summation,
indices $a$ are never lowered and indices $i$ are never raised)%

\begin{align}
L_{\varphi}  &  =-\partial_{\mu}\varphi_{i}^{\ast}\partial^{\mu}\varphi
_{i},\label{free scalar}\\
L_{\psi}  &  =-i\bar{\psi}^{a}\bar{\sigma}^{\mu}\partial_{\mu}\psi^{a},
\label{free spinor}%
\end{align}
is manifestly invariant under a permutation operation $1\rightleftharpoons2$.
So the symmetric group $S_{2}$ \cite{5} defines a discrete symmetry of these
models. The action of the identity operator $s^{1}$ and the transposition
operator $s^{2}$ can be written as%
\begin{align}
s^{1}\varphi_{i}  &  =\delta_{ik}\varphi_{k},\quad s^{1}\psi^{a}=\delta
^{ab}\psi^{b},\nonumber\\
s^{2}\varphi_{i}  &  =\eta_{ik}\varphi_{k},\quad s^{2}\psi^{a}=\eta^{ab}%
\psi^{b}, \label{def permut}%
\end{align}
where $\delta^{ab}=\left\{  1\text{ for }a=b\text{ and }0\text{ for }a\neq
b\right\}  $, $\delta_{ik}=\left\{  1\text{ for }i=k\text{ and }0\text{ for
}i\neq k\right\}  $, $\eta^{ab}=\left\{  0\text{ for }a=b\text{ and }1\text{
for }a\neq b\right\}  $and $\eta_{ik}=\left\{  0\text{ for }i=k\text{ and
}1\text{ for }i\neq k\right\}  $.

Furthermore, we define the modified translation operator $P_{\mu}^{a}$ as a
successive application of a permutation operator $s^{a}$ defined by
(\ref{def permut}) and the four dimensional translation operator $P_{\mu}$%
\begin{equation}
P_{\mu}^{a}=s^{a}P_{\mu},\quad a=1,2\text{ and }\mu=0,1,2,3.
\label{mofied translation}%
\end{equation}
$P_{\mu}^{1}$ is the usual translation (since $s^{1}$ is just the identity)
while $P_{\mu}^{2}$ is the combination of a translation and the transposition
operator. The action of $P_{\mu}^{2}$ is then given by%
\begin{align}
\delta_{\kappa}^{\prime}\varphi_{1}  &  =\kappa^{\mu}\partial_{\mu}\varphi
_{2},\quad\delta_{\kappa}^{\prime}\varphi_{2}=\kappa^{\mu}\partial_{\mu
}\varphi_{1},\label{delta tild1}\\
\delta_{\kappa}^{\prime}\psi^{1}  &  =\kappa^{\mu}\partial_{\mu}\psi^{2}%
,\quad\delta_{\kappa}^{\prime}\psi^{2}=\kappa^{\mu}\partial_{\mu}\psi^{1},
\label{delta tild2}%
\end{align}
where $\kappa^{\mu}$ is an infinitesimal real constant four-vector parameter.
One can easily see that, as it is the case for the usual translation, this
transformation leads also to an invariance of the Lagrangian densities
(\ref{free scalar}) and (\ref{free spinor}). One explicitly finds that
$\delta_{\kappa}^{\prime}L_{\varphi}$ and $\delta_{\kappa}^{\prime}L_{\psi}$
are total derivatives, i.e., $\delta_{\kappa}^{\prime}L_{\varphi}%
=-\partial_{\nu}\left(  \kappa^{\nu}(\partial_{\mu}\varphi_{2}^{\ast}%
\partial^{\mu}\varphi_{1}+\partial_{\mu}\varphi_{1}^{\ast}\partial^{\mu
}\varphi_{2})\right)  $ and $\delta_{\kappa}^{\prime}L_{\psi}=-i\partial_{\nu
}\left(  \kappa^{\nu}(\bar{\psi}^{2}\bar{\sigma}^{\mu}\partial_{\mu}\psi
^{1}+\bar{\psi}^{1}\bar{\sigma}^{\mu}\partial_{\mu}\psi^{2})\right)  $.

\bigskip Moreover, the infinitesimal transformations $\delta^{\prime}$ defined
by (\ref{delta tild1}) and (\ref{delta tild2}) form an abelian algebra. For
two successive transformations $\delta^{\prime}$ of parameters $\kappa$ and
$\zeta$ we get $\delta_{\zeta}^{\prime}\delta_{\kappa}^{\prime}X=\zeta^{\nu
}\kappa^{\mu}\partial_{\nu}\partial_{\mu}X$, where $X$ stands for all the
fields. This leads obviously to%
\begin{equation}
(\delta_{\zeta}^{\prime}\delta_{\kappa}^{\prime}-\delta_{\kappa}^{\prime
}\delta_{\zeta}^{\prime})X=0.
\end{equation}

\bigskip One can also remark that these transformations commute with usual
translations, i.e.,%
\begin{equation}
(\delta_{a}\delta_{\kappa}^{\prime}-\delta_{\kappa}^{\prime}\delta_{a})\psi=0,
\end{equation}
where, as usual, a translation $\delta$ of parameter $a$ is defined by
$\delta_{a}X=a^{\mu}\partial_{\mu}X$.

Finally, it is straightforward to check that the commutator of $\delta
^{\prime}$ with rotations $R$ (i.e., transformations of the Lorentz group)
closes on $\delta^{\prime}$. We have%
\begin{equation}
(R_{\omega}\delta_{\kappa}^{\prime}-\delta_{\kappa}^{\prime}R_{\omega
})X=\delta_{\omega.\kappa}^{\prime}X, \label{commutator with rotations}%
\end{equation}
where $\omega.\kappa\equiv-\omega_{\ \nu}^{\mu}\kappa^{\nu}$ is the
infinitesimal parameter of the resulting $\delta^{\prime}$ transformation. In
deriving (\ref{commutator with rotations}), we used the fact that a rotation
$R$ of infinitesimal parameter $\omega$ acts on any four-vector as $R_{\omega
}V^{\mu}=-\omega_{\ \nu}^{\mu}V^{\nu}$, on any spinor $\psi$ as $R_{\omega
}\psi=-\frac{1}{2}\omega_{\mu\nu}\sigma^{\mu\nu}\psi$ with $\sigma^{\mu\nu
}=\frac{1}{4}(\sigma^{\mu}\bar{\sigma}^{\nu}-\sigma^{\nu}\bar{\sigma}^{\mu})$
and leaves any scalar fields invariant.

Therefore, we can define a modified construction for the extended $N=2$
supersymmetry algebra relying on the nonlinear extension of a Lie algebra. In
this context \cite{3}, the defining commutator contains, in addition to linear
terms, terms that are multilinear in generators, i.e., $\left[  T_{a}%
,T_{b}\right]  =f_{ab}^{c}T_{c}+V_{ab}^{cd}T_{c}T_{d}$ for quadratically
nonlinear algebras. As it was pointed out in \cite{6}, such nonlinear
generalization has also to satisfy Jacobi identities. As extension of the
standard supersymmetry construction where the anti-commutator of two extended
supersymmetry transformations closes on translation, we postulate that it
closes also on the composition of a translation $P_{\mu}$ and a transposition
$s^{2}$, such that%
\begin{equation}
\left\{  Q_{i\alpha},\bar{Q}_{j\dot{\alpha}}\right\}  =2\sigma_{\alpha
\dot{\alpha}}^{\mu}\tau_{ij}^{a}P_{\mu}^{a}, \label{modified susy algebra}%
\end{equation}
where $\tau^{a}=\left(  \tau_{ij}^{a}\right)  $ are the two $2\times2$
matrices given by%
\begin{equation}
\tau^{1}=\left(
\begin{array}
[c]{cc}%
1 & 0\\
0 & 1
\end{array}
\right)  ,\quad\tau^{2}=\left(
\begin{array}
[c]{cc}%
0 & 1\\
1 & 0
\end{array}
\right)  . \label{tau definition}%
\end{equation}
These two matrices form a representation of $S_{2}$ and satisfy the following
relations%
\begin{align}
\tau_{ij}^{a}\tau_{jk}^{b}  &  =\delta^{ab}\delta_{ik}+\eta^{ab}\eta
_{ik},\label{tau property1}\\
\tau_{ij}^{a}\tau_{kl}^{a}  &  =\delta_{il}\delta_{jk}+\eta_{il}\eta_{jk}.
\label{tau property2}%
\end{align}

In view of what precedes on the commutation relations of this modified
translation and the other generators of the Poincar\'{e} algebra (translations
and rotations), the other commutators of the as modified $N=2$ super
Poincar\'{e} algebra read%
\begin{equation}
\left[  P_{\mu}^{a},P_{\nu}^{b}\right]  =0\quad\text{and }\quad\left[
M_{\alpha\beta},P_{\mu}^{a}\right]  =-\eta_{\alpha\mu}P_{\beta}^{a}%
+\eta_{\beta\mu}P_{\alpha}^{a},
\end{equation}
where $M_{\alpha\beta}$ are the generators of the rotations and all other
commutators are identical to those of usual extended $N=2$ super Poincar\'{e}
algebra. Moreover, it is straightforward to check that the modified
supersymmetry algebra (\ref{modified susy algebra}) is consistent\ with all
possible Jacobi identities of the whole algebra.

It is worth noting that $P_{\mu}^{2}$ satisfies, just as the usual
translation, $P_{\mu}^{2}P^{2\mu}=m^{2}$, since permutation operators satisfy
$(s^{a})^{2}=1$ $(a=1,2)$. The Casimir invariant operator $P^{2}$ expressed
thus as $\frac{1}{2}\sum_{a}P_{\mu}^{a}P^{a\mu}=m^{2}$ shows as usual, that
all members of the same multiplet representation are of same masses.

One can also see that the reduction to the simple $N=1$ case leads obviously
to standard results since the permutation operations on a set of one object
are trivial.

We will now show that one can find an off-shell representation of this
algebra, i.e., a multiplet that realizes this modified $N=2$ supersymmetry
algebra \textit{off-shell}.

\section{\bigskip An off-shell representation}

\bigskip We first start with the same field contents that of the double tensor
multiplet (as a generalization of the $N=1$ supersymmetric multiplet of the
gauge spinor superfield \cite{7}). We will show that such a multiplet forms a
representation of the above introduced modified $N=2$ supersymmetric algebra
(\ref{modified susy algebra}) and moreover, a consistent off-shell
construction can be performed. This multiplet contains two Weyl fermions
$\psi$ and $\chi$, two real scalar fields $\varphi_{i},i=1,2$ and two real
2-form gauge potentials $B_{i\mu\nu},\mu(\nu)=0,1,2,3$ and $i=1,2$. All
conventions and notations are the same as in the previous section. In what
follows we work in the two component formalism and adopt the standard
conventions of Wess and Bagger \cite{8}. The Lagrangian density of this
multiplet reads%
\begin{equation}
L=-i\bar{\psi}\bar{\sigma}^{\mu}\partial_{\mu}\psi-i\bar{\chi}\bar{\sigma
}^{\mu}\partial_{\mu}\chi-\frac{1}{2}\partial_{\mu}\varphi_{i}\partial^{\mu
}\varphi_{i}+\frac{1}{2}H_{i\mu}H_{i}^{\mu}, \label{double tensor lagrangian}%
\end{equation}
where $H_{i\mu}$ are the Hodge-duals of the field strengths of the 2-form
gauge potentials, i.e.
\begin{equation}
H_{i}^{\mu}=\frac{1}{2}\varepsilon^{\mu\nu\rho\sigma}\partial_{\nu}%
B_{i\rho\sigma}, \label{hodge dual}%
\end{equation}
with $\varepsilon^{\mu\nu\rho\tau}$ ($\varepsilon^{0123}=+1$) being the
four-dimensional Levi-Civita tensor.

To see that this is indeed a representation of the modified $N=2$
supersymmetric algebra defined by (\ref{modified susy algebra}), we first
check that (\ref{double tensor lagrangian}) is invariant, up to total
derivatives, under the following \textit{modified} extended $N=2$
supersymmetric transformations%
\begin{align}
\delta\psi &  =i\sigma^{\mu}\bar{\xi}_{1}\partial_{\mu}\varphi_{1}%
+i\sigma^{\mu}\bar{\xi}_{2}\partial_{\mu}\varphi_{2}+\sigma^{\mu}\bar{\xi}%
_{1}H_{1\mu}+\sigma^{\mu}\bar{\xi}_{2}H_{2\mu},\\
\delta\chi &  =i\sigma^{\mu}\bar{\xi}_{1}\partial_{\mu}\varphi_{2}%
+i\sigma^{\mu}\bar{\xi}_{2}\partial_{\mu}\varphi_{1}+\sigma^{\mu}\bar{\xi}%
_{1}H_{2\mu}+\sigma^{\mu}\bar{\xi}_{2}H_{1\mu},\\
\delta\varphi_{1}  &  =\xi_{1}\psi+\xi_{2}\chi+h.c.,\\
\delta\varphi_{2}  &  =\xi_{1}\chi+\xi_{2}\psi+h.c.,\\
\delta H_{1}^{\mu}  &  =2i\xi_{1}\sigma^{\mu\nu}\partial_{\nu}\psi+2i\xi
_{2}\sigma^{\mu\nu}\partial_{\nu}\chi+h.c.,\\
\delta H_{2}^{\mu}  &  =2i\xi_{1}\sigma^{\mu\nu}\partial_{\nu}\chi+2i\xi
_{2}\sigma^{\mu\nu}\partial_{\nu}\psi+h.c.
\end{align}
Then recasting the spinor fields $\psi$ and $\chi$ as defined in previous
section, such that $\psi^{1}=\psi$ and $\psi^{2}=\chi$, one can easily write
the above transformations as\bigskip%
\begin{align}
\delta\psi^{a}  &  =i\sigma^{\mu}\bar{\xi}_{i}\tau_{ij}^{a}\partial_{\mu
}\varphi_{j}+\sigma^{\mu}\bar{\xi}_{i}\tau_{ij}^{a}H_{j\mu}%
,\label{delta tensor on psi}\\
\delta\varphi_{i}  &  =\tau_{ij}^{a}\xi_{j}\psi^{a}+\tau_{ij}^{a}\bar{\xi}%
_{j}\bar{\psi}^{a},\label{delta tensor on phi}\\
\delta H_{i}^{\mu}  &  =2i\tau_{ij}^{a}\xi_{j}\sigma^{\mu\nu}\partial_{\nu
}\psi^{a}-2i\tau_{ij}^{a}\bar{\xi}_{j}\bar{\sigma}^{\mu\nu}\partial_{\nu}%
\bar{\psi}^{a}. \label{delta tensor on H}%
\end{align}
A direct computation leads explicitly to $\delta L=-\partial_{\mu}(\psi
^{a}\sigma^{\mu}\bar{\sigma}^{\nu}\xi_{i}\tau_{ij}^{a}\partial_{\nu}%
\varphi_{j}+i\bar{\psi}^{a}\bar{\sigma}^{\mu}\sigma^{\nu}\bar{\xi}_{i}%
\tau_{ij}^{a}H_{j\nu})$ - $\partial_{\mu}(\tau_{ij}^{a}\xi_{j}\psi^{a}%
\partial^{\mu}\varphi_{i}-iH_{i}^{\mu}\tau_{ij}^{a}\xi_{j}\psi^{a}+h.c.)$. We
are now able to compute the action of the commutator of two successive
modified $N=2$ supersymmetric transformations of parameters $\xi$ $(\xi
_{1},\xi_{2})$ and $\zeta$ $(\zeta_{1},\zeta_{2})$ on each field of the
multiplet. Starting with the scalar fields $\varphi_{i}$, we first get%
\begin{align}
\delta_{\zeta}\delta_{\xi}\varphi_{i}  &  =i(\xi_{k}\sigma^{\mu}\bar{\zeta
}_{k}+\bar{\xi}_{k}\bar{\sigma}^{\mu}\zeta_{k})\partial_{\mu}\varphi_{i}%
+i(\xi_{k}\sigma^{\mu}\eta_{kl}\bar{\zeta}_{l}+\bar{\xi}_{k}\bar{\sigma}^{\mu
}\eta_{kl}\zeta_{l})\eta_{ij}\partial_{\mu}\varphi_{j}\nonumber\\
&  +(\xi_{k}\sigma^{\mu}\bar{\zeta}_{k}-\bar{\xi}_{k}\bar{\sigma}^{\mu}%
\zeta_{k})H_{i\mu}+i(\xi_{k}\sigma^{\mu}\eta_{kl}\bar{\zeta}_{l}-\bar{\xi}%
_{k}\bar{\sigma}^{\mu}\eta_{kl}\zeta_{l})\eta_{ij}H_{j\mu}.
\end{align}
Using $\bar{\xi}_{i}\bar{\sigma}^{\nu}\zeta_{j}=-\zeta_{j}\sigma^{\nu}\bar
{\xi}_{i}$, we see that the terms proportional to $\partial_{\nu}\varphi$ are
antisymmetric under the substitution $\bar{\xi}\rightleftharpoons\bar{\zeta}$
such that they are doubled in the commutator $(\delta_{\zeta}\delta_{\xi
}-\delta_{\xi}\delta_{\zeta})\varphi_{i}$, while the terms proportional to $H$
are symmetric under the same substitution, thus they disappear when computing
this commutator. Explicitly, we get%
\begin{equation}
(\delta_{\zeta}\delta_{\xi}-\delta_{\xi}\delta_{\zeta})\varphi_{i}%
=-2i(\zeta_{k}\sigma^{\mu}\bar{\xi}_{k}-\xi_{k}\sigma^{\mu}\bar{\zeta}%
_{k})\partial_{\mu}\varphi_{i}-2i(\zeta_{k}\sigma^{\mu}\eta_{kl}\bar{\xi}%
_{l}-\xi_{k}\sigma^{\mu}\eta_{kl}\bar{\zeta}_{l})\eta_{ij}\partial_{\mu
}\varphi_{j},
\end{equation}
which in regard to (\ref{modified susy algebra}), shows that $(\delta_{\zeta
}\delta_{\xi}-\delta_{\xi}\delta_{\zeta})$ on the scalar fields $\varphi_{i}$
closes off-shell. We turn now to compute the commutator on the spinor fields.
A direct evaluation of $\delta_{\zeta}\delta_{\xi}\psi^{a}$ shows that the
terms in equations of motion of $\psi^{a}$ cancel due to the contribution of
the variation of $H_{i\mu}$, we then obtain%
\begin{align}
\delta_{\zeta}\delta_{\xi}\psi_{\alpha}^{a}  &  =-2i\zeta_{k}\sigma^{\mu}%
\bar{\xi}_{k}\partial_{\mu}\psi_{\alpha}^{a}-2i\zeta_{k}\sigma^{\mu}\eta
_{kl}\bar{\xi}_{l}\eta^{ab}\partial_{\mu}\psi_{\alpha}^{b}\nonumber\\
&  -2i\bar{\zeta}_{k}\bar{\xi}_{k}\sigma_{\alpha\dot{\alpha}}^{\mu}%
\partial_{\mu}\bar{\psi}^{a\dot{\alpha}}-2i\bar{\zeta}_{k}\eta_{kl}\bar{\xi
}_{l}\sigma_{\alpha\dot{\alpha}}^{\mu}\eta^{ab}\partial_{\mu}\bar{\psi}%
^{b\dot{\alpha}},
\end{align}
where we used the identities $(\sigma^{\mu}\bar{\sigma}^{\nu}+\sigma^{\nu}%
\bar{\sigma}^{\mu})_{\alpha}^{\beta}=-2\eta^{\mu\nu}\delta_{\alpha}^{\beta}$,
$\sigma_{\alpha\dot{\alpha}}^{\nu}\bar{\sigma}_{\nu}^{\dot{\beta}\beta
}=-2\delta_{\alpha}^{\beta}\delta_{\dot{\alpha}}^{\dot{\beta}}$ and the
definition $\bar{\sigma}^{\mu\dot{\alpha}\alpha}=\varepsilon^{\dot{\alpha}%
\dot{\beta}}\varepsilon^{\alpha\beta}\sigma_{\beta\dot{\beta}}^{\mu}$ with
$\varepsilon_{12}=\varepsilon^{21}=-1$ is the two-dimensional Levi-Civita
tensor. Noticing that the factors of the terms involving the equations of
motion of $\bar{\psi}^{a}$ are symmetric under the substitution $\bar{\xi
}\rightleftharpoons\bar{\zeta}$ we end up with the following commutator%
\begin{equation}
(\delta_{\zeta}\delta_{\xi}-\delta_{\xi}\delta_{\zeta})\psi^{a}=-2i(\zeta
_{k}\sigma^{\mu}\bar{\xi}_{k}-\xi_{k}\sigma^{\mu}\bar{\zeta}_{k})\partial
_{\mu}\psi^{a}-2i(\zeta_{k}\sigma^{\mu}\eta_{kl}\bar{\xi}_{l}-\xi_{k}%
\sigma^{\mu}\eta_{kl}\bar{\zeta}_{l})\eta^{ab}\partial_{\mu}\psi_{\alpha}^{b},
\end{equation}
which closes off-shell.

Finally, we check the closure on the fields $H_{i}^{\mu}$. Using the identity
$\bar{\xi}_{i}\bar{\sigma}^{\mu\nu}\bar{\sigma}^{\rho}\zeta_{j}=\zeta
_{j}\sigma^{\rho}\bar{\sigma}^{\mu\nu}\bar{\xi}_{i}$ and rearranging terms, we
first find%
\begin{align}
\delta_{\zeta}\delta_{\xi}H_{i}^{\mu}  &  =-2(\xi_{k}\sigma^{\mu\nu}%
\sigma^{\rho}\bar{\zeta}_{k}-\zeta_{k}\sigma^{\rho}\bar{\sigma}^{\mu\nu}%
\bar{\xi}_{k})\partial_{\nu}\partial_{\rho}\varphi_{i}-2(\xi_{k}\sigma^{\mu
\nu}\sigma^{\rho}\eta_{kl}\bar{\zeta}_{l}-\zeta_{k}\sigma^{\rho}\bar{\sigma
}^{\mu\nu}\eta_{kl}\bar{\xi}_{l})\eta_{ij}\partial_{\nu}\partial_{\rho}%
\varphi_{j}\nonumber\\
&  +2i(\xi_{k}\sigma^{\mu\nu}\sigma^{\rho}\bar{\zeta}_{k}+\zeta_{k}%
\sigma^{\rho}\bar{\sigma}^{\mu\nu}\bar{\xi}_{k})\partial_{\nu}H_{i\rho}%
+2i(\xi_{k}\sigma^{\mu\nu}\sigma^{\rho}\eta_{kl}\bar{\zeta}_{l}-\zeta
_{k}\sigma^{\rho}\bar{\sigma}^{\mu\nu}\eta_{kl}\bar{\xi}_{l})\eta_{ij}%
\partial_{\nu}H_{j\rho}.
\end{align}

\bigskip When evaluating the commutator $(\delta_{\eta}\delta_{\xi}%
-\delta_{\xi}\delta_{\eta})H_{i}^{\mu}$, we can see that all terms
proportional to $\partial_{\nu}\partial_{\rho}\varphi$ are of type $\xi
_{k}(\sigma^{\mu\nu}\sigma^{\rho}+\sigma^{\rho}\bar{\sigma}^{\mu\nu}%
)\bar{\zeta}_{l}-(\bar{\xi}\rightleftharpoons\bar{\zeta})$ which is identical
to $i\varepsilon^{\mu\nu\rho\tau}\xi_{k}\sigma_{\tau}\bar{\zeta}_{l}$ so that
all $\varphi$ contributions in the resulting commutator vanish. At the same
time, $\partial_{\nu}H_{i\rho}$ contributions involve terms of type $\xi
_{k}(\sigma^{\mu\nu}\sigma^{\rho}-\sigma^{\rho}\bar{\sigma}^{\mu\nu}%
)\bar{\zeta}_{l}-(\bar{\xi}\rightleftharpoons\bar{\zeta})$ which is identical
to $(\eta^{\mu\rho}\xi_{k}\sigma^{\nu}\bar{\zeta}_{l}-\eta^{\nu\rho}\xi
_{k}\sigma^{\mu}\bar{\zeta}_{l})-(\bar{\xi}\rightleftharpoons\bar{\zeta})$ so
that we end up with the following commutator%
\begin{equation}
(\delta_{\zeta}\delta_{\xi}-\delta_{\xi}\delta_{\zeta})H_{i}^{\mu}%
=-2i(\zeta_{k}\sigma^{\nu}\bar{\xi}_{k}-\xi_{k}\sigma^{\nu}\bar{\zeta}%
_{k})\partial_{\nu}H_{i}^{\mu}-2i(\zeta_{k}\sigma^{\nu}\eta_{kl}\bar{\xi}%
_{l}-\xi_{k}\sigma^{\nu}\eta_{kl}\bar{\zeta}_{l})\eta_{ij}\partial_{\nu}%
H_{j}^{\mu}, \label{algebra on H}%
\end{equation}
where the identities $\sigma^{\mu}\bar{\sigma}^{\nu}\sigma^{\rho}-\sigma
^{\rho}\bar{\sigma}^{\nu}\sigma^{\mu}=2i\varepsilon^{\mu\nu\rho\tau}%
\sigma_{\tau}$ , $\sigma^{\mu}\bar{\sigma}^{\nu}\sigma^{\rho}+\sigma^{\rho
}\bar{\sigma}^{\nu}\sigma^{\mu}=2(\eta^{\mu\rho}\sigma^{\nu}-\eta^{\nu\rho
}\sigma^{\mu}-\eta^{\mu\nu}\sigma^{\rho})$ and $\bar{\zeta}_{k}\bar{\sigma
}^{\mu}\sigma^{\nu}\bar{\sigma}^{\rho}\xi_{l}=-\xi_{l}\sigma^{\rho}\bar
{\sigma}^{\nu}\sigma^{\mu}\bar{\zeta}_{k}$ are used as well as the identity
$\partial_{\nu}H_{i}^{\nu}=0$ which follows from the definition
(\ref{hodge dual}). This ends the proof that the modified $N=2$ supersymmetric
transformations (\ref{delta tensor on psi})-(\ref{delta tensor on H}) form a
supersymmetric algebra that closes off-shell. The $N=2$ multiplet ($\psi^{a}$,
$\varphi_{i}$, $H_{i}^{\mu}$) is then an off-shell multiplet of the modified
$N=2$ supersymmetric algebra defined by (\ref{modified susy algebra}).

It is well known \cite{2} that the double tensor multiplet model has also a
special gauge invariance. Similarly for the above introduced multiplet, it is
easy to check that the Lagrangian density (\ref{double tensor lagrangian}) is
also invariant upon the gauge transformation%
\begin{equation}
\delta_{\Lambda}B_{i}^{\mu\nu}=\partial^{\mu}\Lambda_{i}^{\nu}-\partial^{\nu
}\Lambda_{i}^{\mu},
\end{equation}
where $\Lambda_{i}^{\mu}$ are the spacetime dependent gauge parameters. Since,
by construction (\ref{hodge dual}), the Hodge-duals $H_{i}^{\mu}$ are
obviously invariants upon such a transformation, we have, to make this gauge
transformation appear, to replace $H_{i}^{\mu}$ by the corresponding gauge
potentials $B_{i}^{\mu\nu}$ within the transformations
(\ref{delta tensor on psi})-(\ref{delta tensor on H}). We find%
\begin{align}
\delta\psi^{a} &  =i\sigma^{\mu}\bar{\xi}_{i}\tau_{ij}^{a}\partial_{\mu
}\varphi_{j}+\frac{1}{2}\varepsilon^{\mu\nu\rho\sigma}\sigma_{\mu}\bar{\xi
}_{i}\tau_{ij}^{a}\partial_{\nu}B_{j\rho\sigma},\\
\delta B_{i}^{\mu\nu} &  =-2\tau_{ij}^{a}\xi_{j}\sigma^{\mu\nu}\psi^{a}%
-2\tau_{ij}^{a}\bar{\xi}_{j}\bar{\sigma}^{\mu\nu}\bar{\psi}^{a},
\end{align}
while the transformations of the scalar fields (\ref{delta tensor on phi})
remains the same. We now show that the commutator $(\delta_{\zeta}\delta_{\xi
}-\delta_{\xi}\delta_{\zeta})$ on the gauge potentials $B_{i}^{\mu\nu}$
closes, as previously, off-shell on the combination of translation and
permutations but also on the above defined gauge transformation. After a
similar computation to (\ref{algebra on H}), we find%
\begin{align}
(\delta_{\zeta}\delta_{\xi}-\delta_{\xi}\delta_{\zeta})B_{i}^{\mu\nu} &
=-2i(\zeta_{k}\sigma^{\lambda}\bar{\xi}_{k}-\xi_{k}\sigma^{\lambda}\bar{\zeta
}_{k})\partial_{\lambda}B_{i}^{\mu\nu}-2i(\zeta_{k}\sigma^{\lambda}\eta
_{kl}\bar{\xi}_{l}-\xi_{k}\sigma^{\lambda}\eta_{kl}\bar{\zeta}_{l})\eta
_{ij}\partial_{\lambda}B_{j}^{\mu\nu}\nonumber\\
&  +\partial^{\mu}\Lambda_{i}^{\nu}-\partial^{\nu}\Lambda_{i}^{\mu
},\label{algebra on B}%
\end{align}
where the gauge parameters $\Lambda_{i}^{\mu}$ are given by%
\begin{equation}
\Lambda_{i}^{\mu}=2i\Lambda_{ijk}^{\mu\nu}(\zeta_{j}\sigma_{\nu}\bar{\xi}%
_{k}-\xi_{j}\sigma_{\nu}\bar{\zeta}_{k}),\quad\Lambda_{ijk}^{\mu\nu}=\tau
_{ij}^{a}\tau_{kl}^{a}(\eta^{\mu\nu}\varphi_{l}-B_{l}^{\mu\nu}%
).\label{gauge parameters}%
\end{equation}
In deriving (\ref{algebra on B}) the identity $\varepsilon_{\kappa\tau
\rho\sigma}\varepsilon^{\kappa\mu\nu\lambda}=-[\delta_{\tau}^{\mu}%
(\delta_{\rho}^{\nu}\delta_{\sigma}^{\lambda}-\delta_{\sigma}^{\nu}%
\delta_{\rho}^{\lambda})-\delta_{\tau}^{\nu}(\delta_{\rho}^{\mu}\delta
_{\sigma}^{\lambda}-\delta_{\sigma}^{\mu}\delta_{\rho}^{\lambda})+\delta
_{\tau}^{\lambda}(\delta_{\rho}^{\mu}\delta_{\sigma}^{\nu}-\delta_{\sigma
}^{\mu}\delta_{\rho}^{\nu})]$ is used. As it is generally the case in
supersymmetric gauge theories, these gauge parameters are field dependent. It
is worth noting that comparatively to the standard approach \cite{2} of the
double tensor multiplet, in addition to the fact that in the context presented
here the off-shell construction is possible, the obtained gauge parameters
(\ref{gauge parameters}) do not involve explicitly the spacetime coordinates.

Even if the structure of the modified algebra (\ref{modified susy algebra})
differs from the usual one, we can, in view of the off-shell closure obtained
above, consider the construction of a nilpotent BRST operator. Indeed,
starting from the modified $N=2$ supersymmetry transformations
(\ref{delta tensor on psi})-(\ref{delta tensor on H}), and upon the usual
replacement of the symmetry parameters by the corresponding ghost fields of
opposite statistics, the corresponding BRST construction follows naturally.
Defining the BRST operator $\Delta$ on the fields $\psi^{a}$, $\varphi_{i}$,
$B_{i}^{\mu\nu}$ as%
\begin{align}
\Delta\psi^{a}  &  =i\sigma^{\mu}\bar{\xi}_{i}\tau_{ij}^{a}\partial_{\mu
}\varphi_{j}+\frac{1}{2}\varepsilon^{\mu\nu\rho\sigma}\sigma_{\mu}\bar{\xi
}_{i}\tau_{ij}^{a}\partial_{\nu}B_{j\rho\sigma}+c^{\rho}\partial_{\rho}%
\psi^{a}+\kappa^{\rho}\eta^{ab}\partial_{\rho}\psi^{b},\\
\Delta\varphi_{i}  &  =\tau_{ij}^{a}\xi_{j}\psi^{a}+\tau_{ij}^{a}\bar{\xi}%
_{j}\bar{\psi}^{a}+c^{\rho}\partial_{\rho}\varphi_{i}+\kappa^{\rho}\eta
_{ij}\partial_{\rho}\varphi_{j},\\
\Delta B_{i}^{\mu\nu}  &  =-2\tau_{ij}^{a}\xi_{j}\sigma^{\mu\nu}\psi^{a}%
-2\tau_{ij}^{a}\bar{\xi}_{j}\bar{\sigma}^{\mu\nu}\bar{\psi}^{a}+c^{\rho
}\partial_{\rho}B_{i}^{\mu\nu}+\kappa^{\rho}\eta_{ij}\partial_{\rho}B_{j}%
^{\mu\nu}\\
&  +\partial^{\mu}\Lambda_{i}^{\nu}-\partial^{\nu}\Lambda_{i}^{\mu},
\end{align}
and on the ghosts fields $\xi_{i}$ ($\bar{\xi}_{i}$), $c^{\mu}$, $\kappa^{\mu
}$ and $\Lambda_{i}^{\mu}$ as%
\begin{align}
\Delta\xi &  =0,\\
\Delta c^{\mu}  &  =-2i(\xi_{k}\sigma^{\mu}\bar{\xi}_{k}),\\
\Delta\kappa^{\mu}  &  =-2i(\xi_{k}\sigma^{\mu}\eta_{kl}\bar{\xi}_{l}),\\
\Delta\Lambda_{i}^{\mu}  &  =2i\tau_{ij}^{a}\tau_{kl}^{a}(\eta^{\mu\nu}%
\varphi_{j}-B_{j}^{\mu\nu})(\xi_{k}\sigma_{\nu}\bar{\xi}_{l}),
\end{align}
it is straightforward to show its off-shell nilpotency, i.e., $\Delta^{2}X=0,$
$\forall X$.

\section{Conclusion}

\bigskip\bigskip The main result of this work is that an alternative $N=2$
extension of standard supersymmetry is possible. This is done in the context
of nonlinear extensions of standard Lie algebra by a suitable introduction of
the symmetric group $S_{2}$. The additional nonlinear term being a composition
of a translation and a transposition. The obtained algebra being a
quadratically nonlinear extension of the standard $N=2$ supersymmetric algebra
is however a non usual construction. Indeed, this latter contains structurally
the permutation transformations which are obviously non-local, while, in
deriving the general realization of supersymmetry algebra, only continuous
groups (in particular Lie groups) are usually considered (see e.g. \cite{8}).
This kind of construction will be analyzed in details elsewhere.

The presented result is different from the standard extended $N=2$
supersymmetry, i. e., the anticommutator of two \textit{modified}
supersymmetric transformations must close (at least on-shell) on a mix of
translation \textit{and} permutations, but leads to a consistent algebraic
construction. The reduction \ to the $N=1$ case leads to usual supersymmetry
due to the triviality of the group $S_{1}$ which contains only the identity.
Such a modified extended $N=2$ supersymmetric algebra
(\ref{modified susy algebra}) admits as representation a multiplet that
contains the same fields as the double tensor multiplet (which is in
particular, relevant to type IIB superstring vacua \cite{9}). We have shown
that an off-shell construction is possible, i.e. \textit{without relying on
field equations}. This result has to be compared with the usual double tensor
multiplet for which, in spite of the fact that the bosonic and fermionic
degrees of freedom balance at both on-shell and off-shell levels, the
off-shell construction fails.

Moreover, if a systematic procedure can be considered in order to give the
off-shell version of any given open gauge (local) theory \cite{10}, no such
systematic approach is available in the context of global (rigid) symmetries
such as extended matter supersymmetry, even if specific models exist where the
construction of off-shell realization is possible, i. e., the so-called
$O(2n)$ supermultiplets \cite{11} (see also \cite{12} for a modern review). We
believe that the approach developed here in which the symmetric group $S_{2}$
(or equivalently the group $%
%TCIMACRO{\U{2124} }%
%BeginExpansion
\mathbb{Z}
%EndExpansion
_{2}$) shows up within the nonlinear extension of the usual $N=2$
supersymmetry algebra can offer a new perspective for investigating the
off-shell structure of extended supersymmetric models (e.g. $N=4$
supersymmetric models).


\begin{thebibliography}{99}                                                                                               %


\bibitem {1}B. de Wit, V. Kaplunovsky, J. Louis and D. L\"{u}st, Nucl. Phys. B
\textbf{451}, 53 (1995).

\bibitem {2}F. Brandt, Nucl. Phys. B \textbf{587 }(2000) 543.

\bibitem {3}A. Kent, \textit{"Normal ordered Lie algebras"}. Princeton
preprint IASSNS-HEP-88/04 (1988).

\bibitem {4}L. Frappat, A. Sciarrino and P. Sorba, \textit{"Dictionary on Lie
Algebras and Superalgebras"}, (Academic Press, 2000).

\bibitem {5}H.F. Jones, \textit{"Groups, Representations and Physics"} (IOP
Publishing Ltd. 1998).

\bibitem {6}K. Schoutens, A. Sevrin and P. van Nieuwenhuizen, Commun. Math.
Phys. 124, 87-103 (1989).

\bibitem {7}W. Siegel, Phys. Lett. B85 (1979) 333.

\bibitem {8}J. Wess and J. Bagger, "\textit{Supersymmetry and Supergravity}%
"\ (Princeton, NJ: Princeton University Press, 1983).

\bibitem {9}J. Louis and K. F\"{o}rger, Nucl. Phys. Proc. Suppl. \textbf{55} B
(1997) 33.

\bibitem {10}N. Djeghloul and M. Tahiri, Phys. Rev. D \textbf{66}, 065010 (2002).

\bibitem {11}U. Lindstr\"{o}m and M. Ro\v{c}ek, Commun. Math. Phys.
\textbf{115} (1988) 21.

\bibitem {12}S. M. Kuzenko, J. Phys. A \textbf{43}, 443001 (2010).
\end{thebibliography}
\end{document}